\documentclass[aps,prl,twocolumn,showpacs,superscriptaddress]{revtex4-1}  % for review and submissiongener
\usepackage[latin1]{inputenc}
\usepackage[T1]{fontenc}
\usepackage{graphicx}  % needed for figures
\usepackage{dcolumn}   % needed for some tables
\usepackage{bm}        % for math
\usepackage{amsmath,amssymb,amsfonts}   % for math
\usepackage{psfrag}
\usepackage{epsfig}
\usepackage{array}
\usepackage{color}
\usepackage{hyperref}
\usepackage{multirow}

\newcommand{\Sum}[2]{\underset{#1}{\overset{#2}{\sum}}}

\def\conf{x}
\def\bR{{\conf^\prime}}
\def\p{\pi}
\def\E{\mathbb{E}}
\def\O{{\cal O}}
\begin{document}

\title{Systematic lowering of the scaling of Monte Carlo calculations by partitioning and subsampling}

\author{Antoine Bienvenu}
\affiliation{Laboratoire de Chimie Th\'eorique, Sorbonne Universit\'e and CNRS, F-75005 Paris, France}
\author{Jonas Feldt}
\affiliation{Laboratoire de Chimie Th\'eorique, Sorbonne Universit\'e and CNRS, F-75005 Paris, France}
\author{Julien Toulouse}
\affiliation{Laboratoire de Chimie Th\'eorique, Sorbonne Universit\'e and CNRS, F-75005 Paris, France}
\affiliation{Institut Universitaire de France, F-75005 Paris, France}
\author{Roland Assaraf}
\email{assaraf@lct.jussieu.fr}
\affiliation{Laboratoire de Chimie Th\'eorique, Sorbonne Universit\'e and CNRS, F-75005 Paris, France}

\date{June 30, 2022}

\begin{abstract}
We propose to compute physical properties by Monte Carlo calculations using conditional expectation values. The latter are obtained on top of the usual Monte Carlo sampling by partitioning the physical space in several subspaces or fragments, and subsampling each fragment (i.e., performing side-walks) while freezing the environment. No bias is introduced and a zero-variance principle holds in the limit of separability, i.e. when the fragments are independent. In practice, the usual bottleneck of Monte Carlo calculations -- the scaling of the statistical fluctuations as a function of the number of particles $N$ -- is relieved for extensive observables.
We illustrate the method in variational Monte Carlo on the 2D Hubbard model and on metallic hydrogen chains using Jastrow-Slater wave functions. A factor $\O(N)$ is gained in numerical efficiency.
\end{abstract}

\maketitle

Many domains of physics involve large dimensional integrals which can be computed efficiently with Monte Carlo methods, e.g. statistical physics \cite{Binder:1339249}, quantum physics applied to molecules and solids \cite{Foulkes2001}, or nuclear physics \cite{Lyn2019}.
Monte Carlo methods reinterpret the energy or other properties as the  expectation value of a random variable $O$ over a probability distribution $\pi$ on a configuration space $\Omega$
\begin{equation}
  \E  (O) = \int_{\conf \in \Omega}  O(\conf) \pi(\conf) d\conf.
\label{EO}
\end{equation}
Typically, the configuration $x$ corresponds to the $3N$ coordinates of the particles in physical space, but it can also correspond to the $N$ trajectories of the particles in the path-integral formulation of quantum mechanics. The probability distribution $\pi$ depends on the context. For example, in equilibrium statistical physics, $\pi$ is the Gibbs distribution. In variational Monte Carlo (VMC), $\pi=\Psi^2$ is the probability density of a wave function $\Psi$, and if $O= (H\Psi)/\Psi$ is the local energy for a given Hamiltonian $H$ then $\E(O)$ is the variational energy.
Expectation values are computed using the ergodic theorem which states that the integral can be written as a time average,
$\E(O) = \lim_{M \to \infty} (1/M) \sum_{i=1}^M O(\conf^i)$,
where the sequence of $M$ configurations $(\conf^i)$ is built from a $\pi$-invariant ergodic stochastic process (usually a Markov chain). The sequence $(\conf^i)$ is called a sample of the distribution $\pi$.

The bottleneck of Monte Carlo methods comes from the statistical fluctuations which usually grow with the system size, as measured by the number of particles $N$.
For a sample of sufficiently large  size $M$, the statistical uncertainty $\sigma$ on the estimation of $\E(O)$ is
\begin{equation}
  \sigma =   \sqrt{\frac{V(O)c}{M}},
  \label{sm2}
\end{equation}
where $V(O)=\E(O^2) -\E(O)^2$ is the variance of $O$ and $c>1$ is a correlation factor which takes into account that the configurations are not fully independent.
According to Eq.~(\ref{sm2}), reaching a given precision $\sigma$ requires a CPU time $t_M=M t_1$ proportional to both the time $t_1$ of performing one step of the sampling and to the variance $V(O)$.
The numerical efficiency of the method can then be measured by the asymptotically $M$-independent quantity
\begin{equation}
  \sigma^2 t_M = V(O)c t_1,
  \label{numeff}
\end{equation}
which should  be as small as possible for maximal efficiency. In the present work, we will not be concerned about the correlation factor $c$ which sometimes diverges with $N$ (e.g. near criticality). A large corpus of work is devoted to reducing its scaling as a function of $N$, such as parallel tempering based methods (see, e.g., Refs.~\cite{PhysRevD.40.2035,weare2007}). Equation~(\ref{numeff}) indicates a more crucial double penalty of Monte Carlo methods for large systems: both $t_1$ and $V(O)$ grow with system size $N$.
 This double penalty is for example at the origin of the main bottleneck in computing the VMC energy of a fermionic system in real space \cite{Foulkes2001,TOULOUSE2016285}.
Evaluating the wave function involves indeed calculating a Slater determinant of order $\O(N\times N)$ which costs $t_1=\O(N^3)$ while the variance is typically extensive, $V(O) \propto N$, thus rising the scaling of the overall cost to $\O(N^4)$. This scaling is still larger than some deterministic methods like the celebrated Kohn-Sham density-functional theory which scales as $\O(N^3)$ for a spatially delocalized (i.e., metallic) system \cite{MOHR201864}.

The extensivity of the variance has a physical origin. A large system can in general be approximated by a collection of independent fragments.
This ideal case corresponds to the separability limit where the random variable $O$ is the sum of independent variables $O_k$ on each fragment indexed by $k$, i.e. $O = \sum_k O_k$, and the variance is then $V(O) = \sum_k V(O_k) \propto N$.
It is possible to reduce considerably the variance using an improved estimator $\tilde{O}$ built from the approximate solution of a partial differential equation  \cite{assaraf:4682, Mira2012,Borgis_doi:10.1080/00268976.2013.838316}. But this type of improved estimator is still a sum of independent random variables in the separability limit, i.e. $\tilde{O} = \sum_k \tilde{O}_k$, and thus does not change the scaling with respect to $N$ but only reduces the prefactor \cite{assaraf_PhysRevE.89.033304}.

To reduce the global computational scaling, a common and obvious strategy is to reduce the cost of the sampling. Some distributions $\pi$ can be sampled with a linear-scaling algorithm, i.e. $t_1 = \O(N)$, reducing the overall cost to an ideal scaling $\O(N^2)$. One can for example try to use the sparsity of the Slater matrix when localized Wannier functions are used \cite{PhysRevLett.87.246406}. But such sparsity is highly dependent on the physics of the system, and does not hold for a metallic system. Besides, this linear scaling is only theoretical because of memory-access slow down as $N$ increases.
Another strategy consists in using a stable-versus-chaos stochastic dynamics \cite{PhysRevE.90.063317} but finding such a stochastic dynamic is not straightforward \cite{Assaraf2017}.

Here we propose to reduce the global computational scaling by using the locality of physical observables. The idea of using the locality of information to reduce the variance is not new: the strong locality in time of the Schr\"odinger equation (a first-order partial differential equation in time) has for example been exploited to remove the dynamical sign problem for bosonic systems \cite{Cohen_2015}.
Recently, a method was proposed \cite{doi:10.1021/acs.jctc.0c01069} to exploit the low correlation between different core regions in a molecule, resulting in a reduced scaling as a function of the atomic charge $Z$ but not as a function of $N$.
The present work  exploits the fact that in an extended physical system (including a metallic system) correlations between large fragments are small.
We construct an improved estimator $\tilde{O}$ with a variance having a reduced scaling with respect to $N$, without changing the scaling of $t_1$, therefore achieving a reduction of the overall computational scaling.
The present work shares the same general philosophy as other fragment-based methods (see, e.g., Refs.~\onlinecite{Wite1992,KniCha-PRL-12,Zahariev2021CombinedQM}). However, while the latter methods are systematic techniques to find a good compromise between a smaller computational time and a larger systematic error, in the present method the reduction of the computational scaling is done without introducing any systematic error.

\vspace{0.2cm}
{\it Theory ---}
A configuration of particles is written as $\conf=(\conf_j)_{j\in J}$ where $\conf_j$ is the $j^\text{th}$ coordinate and $J$ is the list of coordinate indexes. For a given configuration $\conf=(\conf_j)_{j\in J}$, we define a partition of $J$ as $p$ disjoint sublists  $J_k(\conf) \subset J$ such that $\bigcup_{k=1}^p J_k(\conf) =J$. We then define $p$ fragments as subsets $\Omega_k(\conf)$ of the configuration space $\Omega$ such that for all
$\bR  \in \Omega_k(\conf)$, (i) $\bR$ differ from $\conf$ only  by the coordinates indexed by $J_k$, and (ii) $\Omega_k(\bR) = \Omega_k(\conf)$. In short $\Omega_k$ can be seen as a parameter which specifies the positions of the frozen particles in the environment of a fragment.
We then introduce the following improved estimator
\begin{equation}
  \tilde{O}    \equiv  O + \sum_{k=1}^p \lambda_k (\E(O|\Omega_k) -  O),
  \label{tildeo}
\end{equation}
where $\lambda_k$ are constants (or more generally functions of $\Omega_k$) 
and  $\E(O | \Omega_k)$ is the conditional expectation value of the random variable $O$ with respect to $\Omega_k$, defined as the random variable obtained by partial averaging of $O$ over only configurations $\bR \in \Omega_k$
\begin{equation}
  \E(O | \Omega_k) \equiv \frac{\int_{{\bR} \in \Omega_k} O(\bR) \p (\bR) d\bR  }{\int_{\bR \in \Omega_k} \p(\bR) d\bR}.
\end{equation}
The estimator $\tilde{O}$ in Eq.~(\ref{tildeo}) is always not biased, 
i.e. $\E(\tilde{O})=\E(O)$. Indeed $\E(O|\Omega_k) -  O$ has a zero expectation value because of the well-known law of total expectation $\E(\E(O|\Omega_k)) = \E(O)$.
This law can be proven starting from Eq.~(\ref{EO}), i.e. $\E(\E(O|\Omega_k)) = \int  \E(O|\Omega_k)\pi(x)dx$, and decomposing the integral over $x$ as an integral over the environment variable $\Omega_k$ and an integral over $x^\prime \in \Omega_k$.
Let us prove now that the estimator $\tilde{O}$ has a zero-variance property in the separability limit when we choose $\lambda_k=1$ $\forall k$. In this limit, $O$ is a sum of $p$ independent contributions on each fragment, $O = \sum_{k=1}^p O_k((\conf_j)_{j\in J_k})$.
Independence implies that $\E(O_k |\Omega_k) = \E(O_k)$ and  $\E(O_l |\Omega_k)=O_l$ if $l \ne k$, therefore
$\E(O | \Omega_k) -O = \E(O_k)-O_k$ and
\begin{equation}
\tilde{O} = \sum_{k=1}^p \E(O_k) = \E(O).
\end{equation}
In this limit $\tilde{O}$ is a constant, only one parent configuration $\conf$ is sufficient for sampling $\tilde{O}$,  the algorithm becomes equivalent to $p$ independent Monte Carlo simulations of the $p$ subsystems.

Of course, we do not know $\E(O|\Omega_k)$, but we can sample it from the marginal distribution $\pi(. | \Omega_k)$. This is done through a side-walk which samples only $\Omega_k$, i.e. moving the coordinates indexed by $J_k$ in a given fragment while the other coordinates are frozen.
From now on we will use the practical definition of the  improved estimator
\begin{equation}
  \tilde{O} \equiv O + \sum_{k=1}^p  \frac{\lambda_k }{m_k}\sum_{i=1}^{m_k} (O_k^i-O),
\label{improvedest}
\end{equation}
where $O_k^i$ is the value of the random variable $O$ at the $i^\text{th}$ step of the $k^\text{th}$ side-walk (moving only the coordinates indexed by $J_k$) of length $m_k$. A direct way to see that the estimator in Eq.~(\ref{improvedest}) is not biased is to note that  $\E(O_k^i-O)=0$ as $O_k^i$ and $O$ share the same distribution $\p$, since the side-walk and the main walk both sample $\p$.
We expect this scheme that we call the partition Monte Carlo (PMC) method to reduce  the variance with a low numerical cost because the $p$ subsamplings correspond to handling $p=\O(N)$ low-dimensional problems.
The practical formula in Eq.~(\ref{improvedest}) is equivalent to the theoretical definition in Eq.~(\ref{tildeo}) in the limit $m_k \to \infty$ thanks to the ergodic theorem.
In practice, the parameters $\lambda_k$ and $m_k$ have to be adjusted to lower the variance of $\tilde{O}$ for a given CPU time. Also, for optimal efficiency, we can generalize the estimator $\tilde{O}$ in Eq.~(\ref{improvedest}) using instead of $O_k^i -O$ the control variate $G_k^i-G_k$ provided it converges to the former in the separability limit. $G_k$ can be obtained from $O$ by neglecting terms outside of the fragment $k$, reducing the computational cost while retaining the unbiasedness and the zero-variance property in the separability limit. For example when computing the variational energy of a molecule, i.e. $O = (H\Psi)/\Psi$, we take $G_k=(H_k\Psi)/\Psi$ where $H_k$ is the truncated Hamiltonian
\begin{equation}
  H_k =  \sum_{i=1}^{n_k} \left( -\frac{1}{2}\nabla_i^2 - \sum_{A} \frac{Z_A}{r_{iA}} + \sum_j \frac{1}{r_{ij}} \right),
  \label{trunch}
\end{equation}
where the index $i$ runs over the $n_k$  electrons in the fragment $k$. The first term is the kinetic-energy operator and the last two terms are the Coulomb interactions of the electrons of the fragment with the nuclei $A$ (charges $Z_A$) and electrons $j$ lying in a given neighborhood of the fragment.

Let us see now how the PMC method relieves the variance bottleneck. As an example, we consider VMC calculations using Jastrow-Slater wave functions
\begin{equation}
\Psi(x) = e^{J(x)} \Phi(x),
\end{equation}
where $J(x)$ is any real symmetric function of the electron configuration $x$, and $\Phi(x) = \det(A)$ with the Slater matrix $A=X C$ where $X$ is a rectangular matrix of localized atomic orbitals (Kronecker functions in the case of a lattice  model) and $C$ is the rectangular matrix of the orbital coefficients.
For one fragment of the system we introduce now the matrix $P$ which selects the lines corresponding to the electrons of that fragment. For a side-walk in that fragment, $X$ takes different values $X^\prime$ such that only the lines $PX$ might differ from the lines $PX^\prime$. The new determinant is~\cite{doi:10.1063/1.4948778,slater_2017}
\begin{eqnarray}
  \Phi(x') &=& \det(X^\prime C)
\nonumber\\
           &=& \det(A) \det(X^\prime C A^{-1})
\nonumber\\
           &=& \det(A) \det (P X^\prime Q^T Q C A ^{-1} P^T),
\label{detlemma}
\end{eqnarray}
where we have used the determinant lemma. We inserted the projector $Q^T Q$ where  $Q^T$ selects on the right of $PX^\prime$ only the few columns which may differ from zero for this fragment. These columns are very few because the atomic orbitals are localized.
In conclusion updating the determinant along the side-walk is equivalent to multiplying it by a low-order effective Slater determinant
\begin{equation}
  \Phi(x') = \det(A) \det (\bar{X} \bar{C}),
\label{lowdet}
\end{equation}
where
$\bar{X} = P X^\prime Q^T$ and $\bar{C} = Q C A ^{-1} P^T$.
The matrix $\bar{C}$ represents effective orbitals for the fragment and is computed only once at each step of the usual main walk, at a $\O(N^3)$ numerical cost. Once $\bar{C}$ has been built and stored the side-walk costs only $\O(n^3)$ where $n$ is the number of electrons in the fragment. The local energy of the subsystem involves a truncated Hamiltonian and can be computed with the same cost $O(n^3)$~\cite{doi:10.1063/1.4948778,slater_2017}. The cost of subsampling $\O(N)$ fragments is thus $\O(N)$ for an extended system with a finite correlation length. This allows us to perform up to  $\sum_k m_k = \O(N^3)$ total steps in the side-walks without modifying the scaling of the main walk. Therefore, we can perform $m_k =\O(N^2)$ steps in each fragment and the improved estimator in Eq.~(\ref{improvedest}) will have consequently a variance reduced by a factor up to $\O(N^2)$, which is achieved in the separability limit.

%%%%%%%%%%%%%%%%%%%%%%%%%%%%%%%%%%%%%%%%%%%%%%%%%%%%%%%
\begin{figure*}[t]
\centering
\includegraphics[scale=0.24]{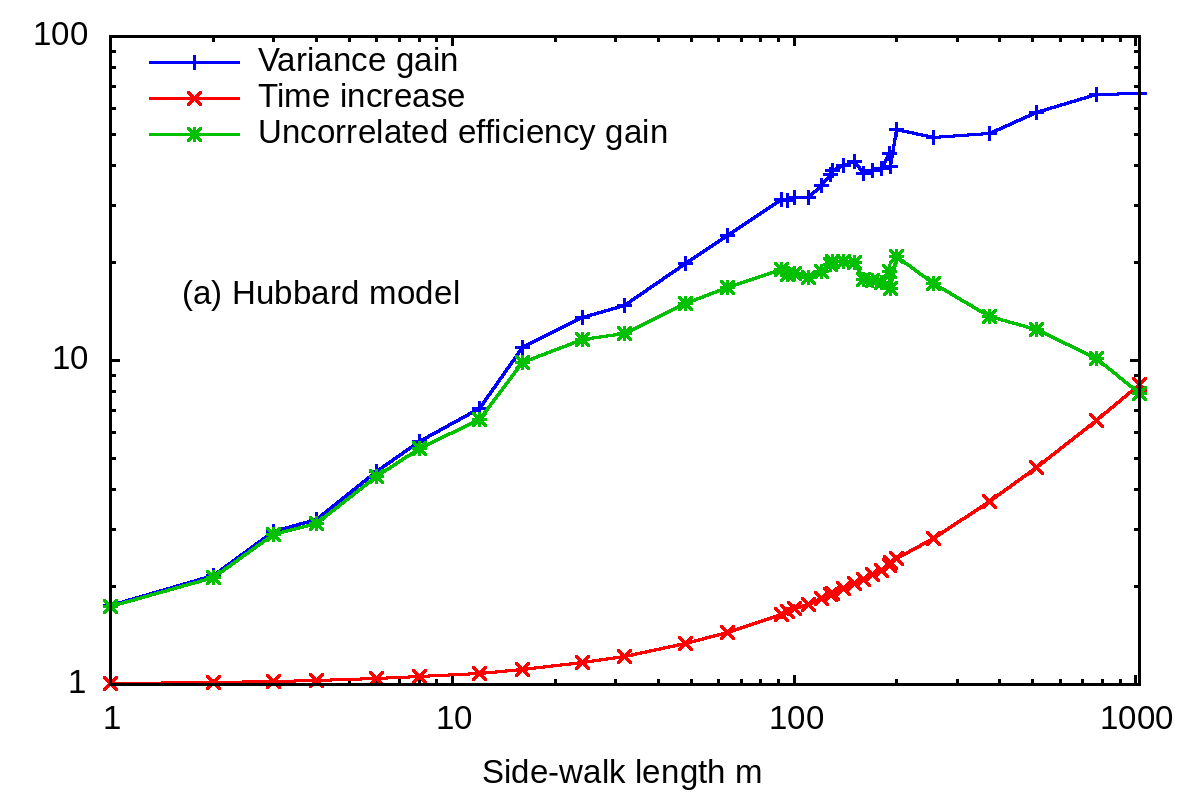}
\includegraphics[scale=0.24]{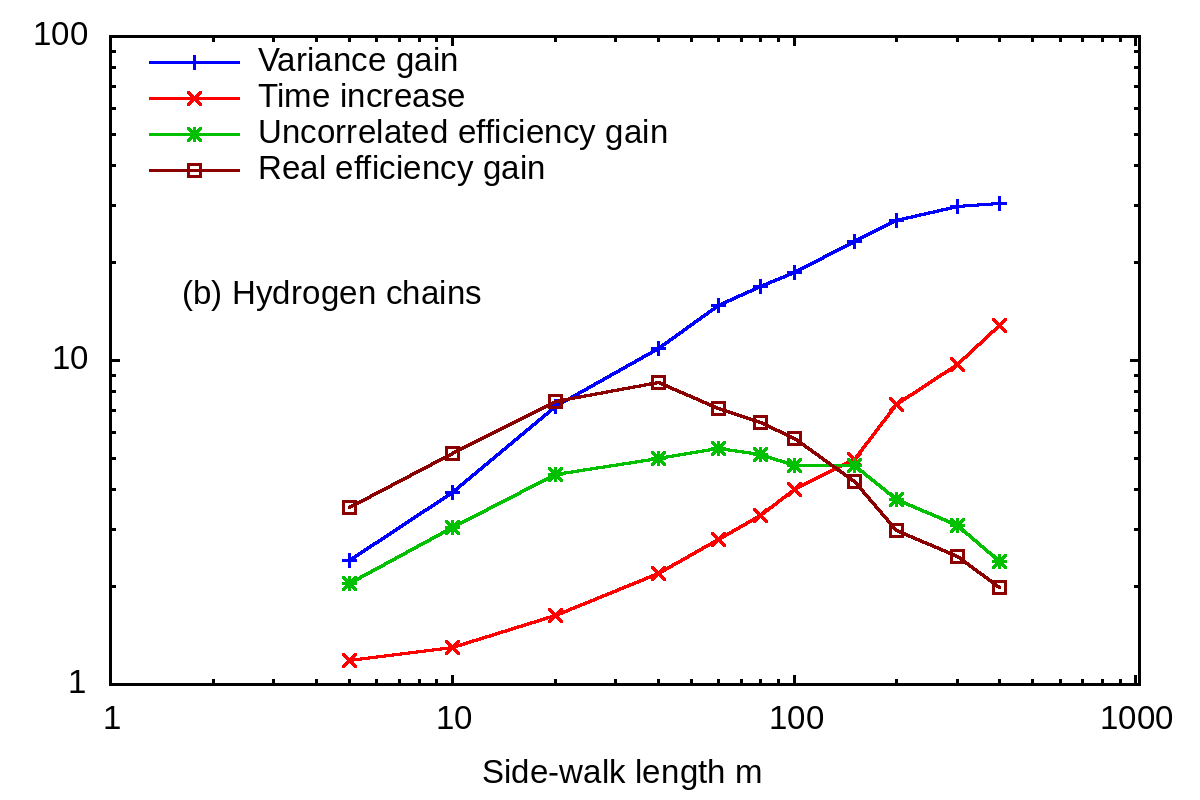}
\caption{Variance gain $V_\text{VMC}/V_\text{PMC}$, time increase $t_\text{PMC}/t_\text{VMC}$,  uncorrelated efficiency gain $(V_\text{VMC}t_\text{VMC})/(V_\text{PMC} t_\text{PMC})$   of the PMC method over the standard VMC method as a function of the side-walk length $m$ for (a) the $20 \times 20$ square Hubbard model (half filling) and (b) the H$_{320}$ metallic hydrogen chain.
The real efficiency gain $(\sigma^2_\text{VMC} t_\text{VMC})/(\sigma^2_\text{PMC} t_\text{PMC})$ differs from the uncorrelated efficiency gain only for hydrogen chains.
}
\label{fig1}
\end{figure*}
%%%%%%%%%%%%%%%%%%%%%%%%%%%%%%%%%%%%%%%%%%%%%%%%%%%%%%%

\vspace{0.2cm}
{\it Results ---} We now illustrate the PMC method on the calculation of the ground-state energy of the 2D Hubbard model and of metallic hydrogen chains.

The Hubbard systems that we employ consist in 2D square grids of $L\times L$ sites with periodic boundary conditions, filled to half-capacity with $N\approx L^2$ electrons evenly distributed between the spins. Designating by $c_{i\sigma}^\dag$ and $c_{i\sigma}$ the creation and annihilation operators of site $i$ with spin $\sigma\in \{\uparrow,\downarrow\}$, and by $n_{i\sigma}=c_{i\sigma}^\dag c_{i\sigma}$ the corresponding number operators, the Hamiltonian takes the form~\cite{cyrot1977hubbard}
\begin{equation}
H= -\Sum{i\neq j,\sigma}{}t_{ij}c^\dag_{i\sigma} c_{j\sigma} + U \Sum{i}{}n_{i\uparrow}n_{i\downarrow},
\end{equation}
where $t_{ij}=1$ if $i$ and $j$ are adjacent, and $t_{ij}=0$ otherwise, and $U=1$ is the on-site interaction parameter. We have chosen the trial ground-state wave function to be a Slater determinant of plane waves without any Jastrow factor. We choose the subsystems as adjacent squares of $l\times l$ sites. The number of iterations of the main walk is kept constant at $M=500$.

As an example of a simple system with a continuum configuration space, we consider metallic hydrogen chains with a regular interatomic distance of $1.4~a_0$. The Hamiltonian is given by Eq.~(\ref{trunch}) except of course that there is no restriction in the sums for the full system.
For the trial ground-state wave function, we use a simple Jastrow function \cite{doi:10.1021/acs.jctc.0c01069} multiplied by the Hartree-Fock Slater determinant obtained from a basis made of the exact hydrogen 1s orbital on each atom.
We choose the subsystems as consisting in $n$ adjacent hydrogen atoms.

The first parameter of the PMC method whose impact is to be explored is the side-walk length $m$ (chosen to be the same for all subsystems). Figure \ref{fig1} reports the variance gain $V_\text{VMC}/V_\text{PMC}$, the CPU time increase $t_\text{PMC}/t_\text{VMC}$, and the uncorrelated efficiency gain $(V_\text{VMC}t_\text{VMC})/(V_\text{PMC} t_\text{PMC})$ [efficiency gain assuming  a correlation factor $c=1$] of the PMC method over the standard variational Monte Carlo (VMC) method.
The efficiency gain is plotted as a function of the side-walk length $m$ for the 2D Hubbard model with total size $L=20$ and subsystem size $l=5$, and for hydrogen chains with $N=320$ total atoms and $n=12$ atoms in the subsystems. Two regimes are clearly visible. For small $m$, the variance gain increases linearly with $m$ while the CPU time is almost constant (the cost of a side-walk step is very small compared to that of a main walk step). This leads to a linear increase of the uncorrelated efficiency gain. For large $m$, the variance gain saturates while the CPU time ratio increases linearly, driving the uncorrelated efficiency gain down. Between these two regimes, there is a plateau corresponding to optimal values of the side-walk length $m$.
The saturation of the variance gain originates from the correlation between subsystems. Indeed, if the subsystems were independent, the variance would converge to zero as $m$ increases (zero-variance principle in the separability limit) and the variance gain to infinity.

One may ask the role of the correlation factor $c$ in Eq.~(\ref{sm2}). For the Hubbard model, $c$ has been found to be very close to $1$ leading to a real efficiency gain almost identical to the uncorrelated efficiency gain.
For the hydrogen chains  $c \simeq 2.5$ for $m=0$ (VMC) and  $c$  is reduced for small $m$  (about $40\%$ less for $H_{320}$ and $m \in [5,40]$) before increasing slowly for larger values of $m$. This explains the difference between the uncorrelated efficiency gain and the real efficiency gain $(\sigma^2_\text{VMC} t_\text{VMC})/(\sigma^2_\text{PMC} t_\text{PMC})$ in Fig.~\ref{fig1}. In particular, the optimal real efficiency gain is $40\%$ higher than the optimal uncorrelated efficiency gain.
%%%%%%%%%%%%%%%%%%%%%%%%%%%%%%%%%%%%%%%%%%%%%%%%%%%%%%%%%%%%%%%%%%%%%%%%%%%%%
\noindent\begin{figure}[t]
\centering
\includegraphics[scale=0.18]{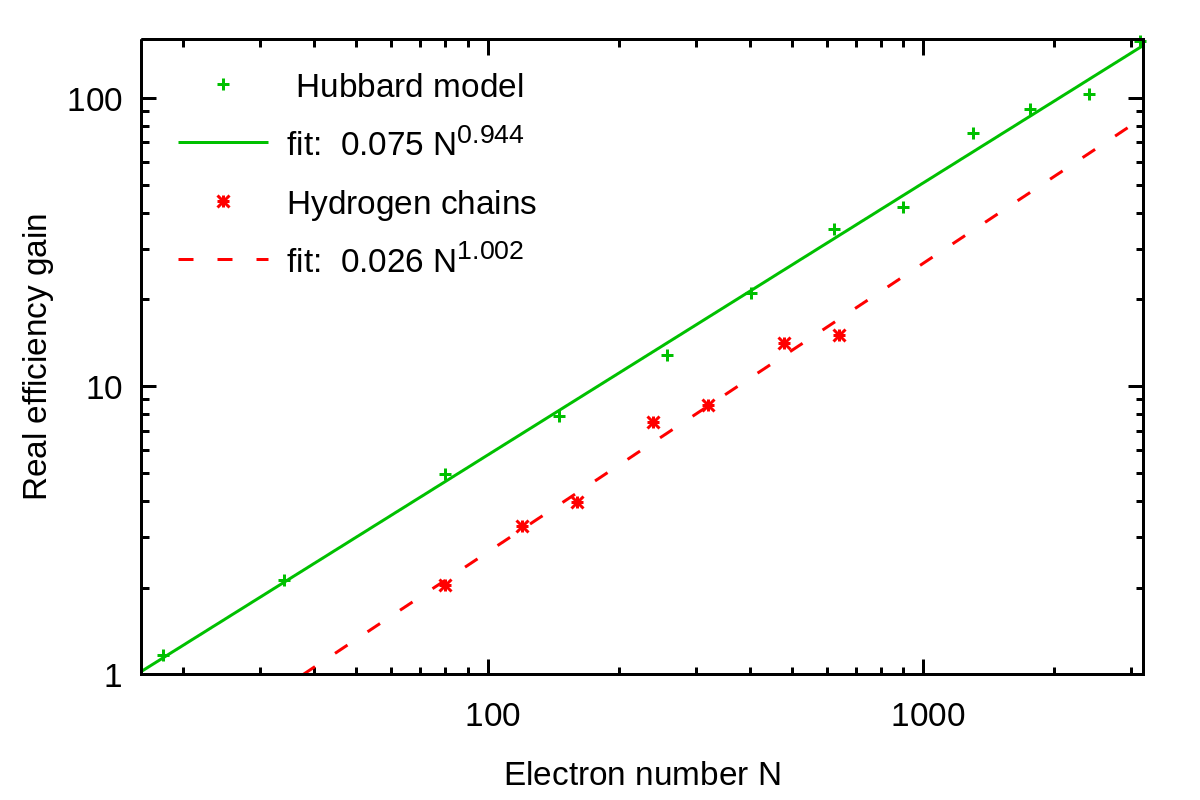}
\caption{Optimal real efficiency gain $(\sigma^2_\text{VMC} t_\text{VMC})/(\sigma^2_\text{PMC} t_\text{PMC})$ as a function of electron number $N$ for the Hubbard model and metallic hydrogen chains.}
\label{fig2}
\end{figure}
%%%%%%%%%%%%%%%%%%%%%%%%%%%%%%%%%%%%%%%%%%%%%%%%%%%%%%%%%%%%%%%%%%%%%%%%%%%%%

%%%%%%%%%%%%%%%%%%%%%%%%%%%%%%%%%%%%%%%%%%%%%%%%%%%%%%%%%%%%%%%%%%%%%%%%%%%%%
\noindent\begin{figure*}[t]
\centering
\includegraphics[scale=0.24]{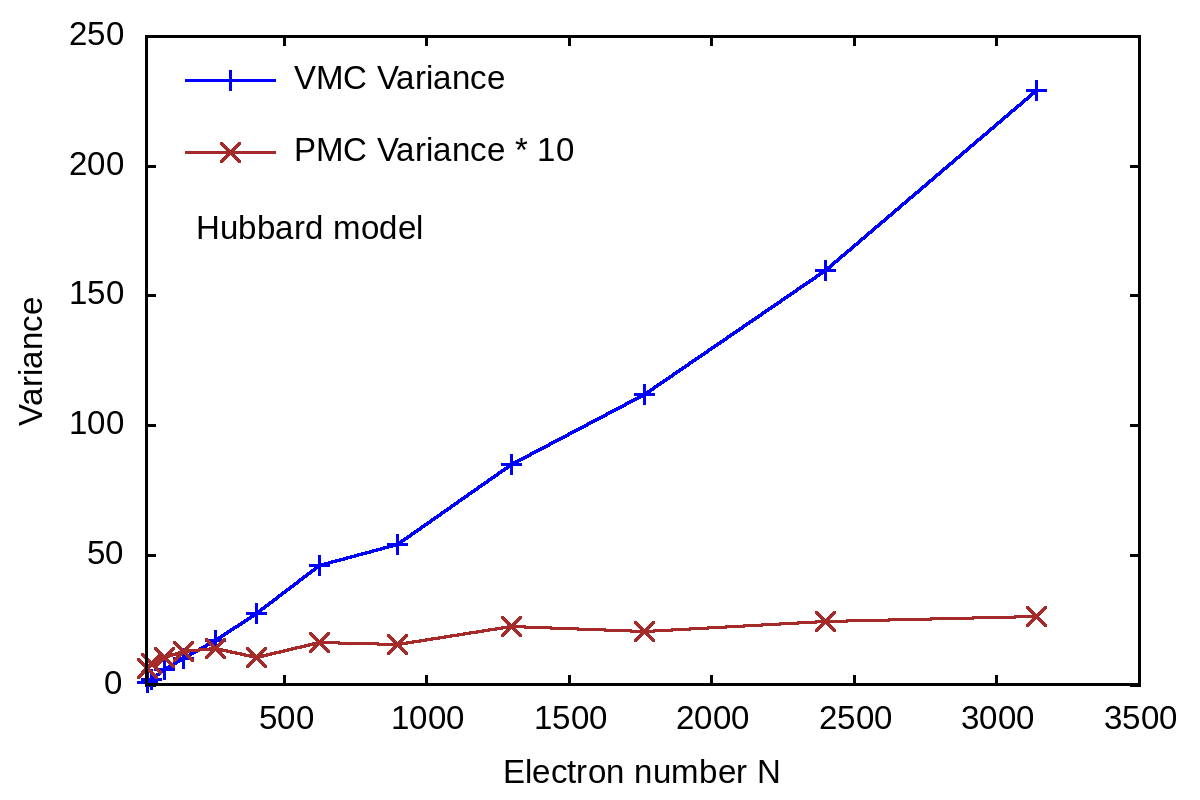}
\includegraphics[scale=0.24]{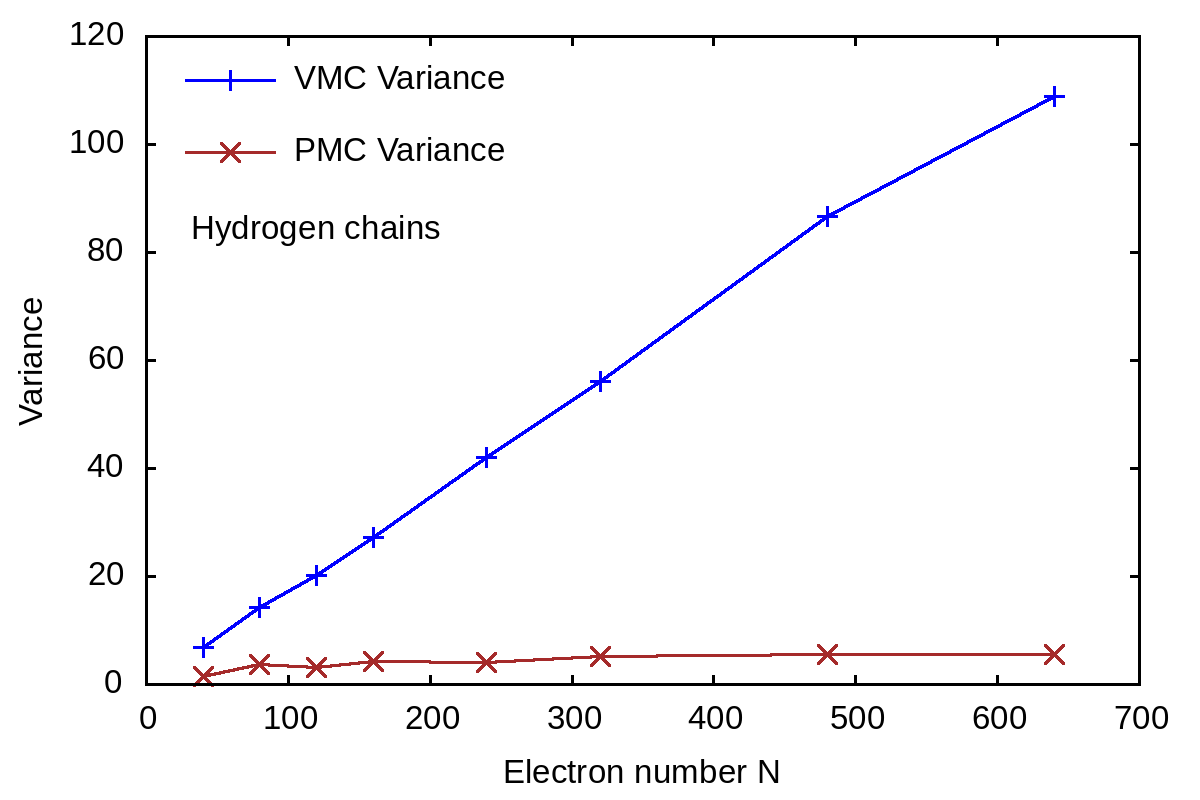}
\caption{Variance of the local energy in standard VMC and in PMC (for optimal $m$) as a function of electron number $N$ for (a) the Hubbard model (PMC variance multiplied by 10 on the plot) and (b) metallic hydrogen chains.}
\label{fig3}
\end{figure*}
%%%%%%%%%%%%%%%%%%%%%%%%%%%%%%%%%%%%%%%%%%%%%%%%%%%%%%%%%%%%%%%%%%%%%%%%%%%%%
We now consider systems of increasing sizes. For the Hubbard model, the optimal subsystem size has been found to be $l \approx \sqrt{L}$, and similarly for the metallic hydrogen chains we find $n \approx \sqrt{N/2}$. The fact that the optimal subsystem size does not saturate to a finite value as the system size increases is an indication of the non-separability of the system.
The optimal side-walk length $m$ also increases with system size since larger systems result in more decorrelated subsystems and cheaper side-walks compared to the main walk. Figure~\ref{fig2} reports the real efficiency gain as a function of the electron number $N$ for the Hubbard model and the hydrogen chains up to $N$ of the order of $10^3$. Both metallic systems present a real efficiency gain scaling linearly with $N$, which hovers around $0.075 N$ for the Hubbard model and $0.025 N$ for the hydrogen chains. This real efficiency gain is entirely achieved by decreasing the variance of the local energy from $\O(N)$ to a behavior close to $\O(1)$, as shown in Fig.~\ref{fig3}. Of course, we have checked that computing $\E(O)$ and $\E(\tilde{O})$ always gives the same answer within the error bars, in agreement with the unbiasedness of $\tilde{O}$.

\vspace{0.2cm}
       {\it Conclusions ---} We introduced a general and simple method to reduce the scaling of Monte Carlo calculations of extensive properties.
       It only requires to have an explicit formula [Eq.~(\ref{EO})] for the integral to be computed, and therefore can be used in any Markov Chain Monte Carlo application.
The method was illustrated on VMC calculations of metallic systems of $N$ particles, providing an efficiency gain of order $\O(N)$.
The present idea can be applied in many contexts, including fixed-node path-integral Monte Carlo approaches~\cite{Baroni1999ReptationQM,Shumway2006} since these schemes sample explicit probability distributions.
Finally, the method can in principle be extended to derivatives of extensive properties to reduce the scaling for calculating response properties or optimizing variational wave functions.


\begin{thebibliography}{0}%
\makeatletter
\providecommand \@ifxundefined [1]{%
 \@ifx{#1\undefined}
}%
\providecommand \@ifnum [1]{%
 \ifnum #1\expandafter \@firstoftwo
 \else \expandafter \@secondoftwo
 \fi
}%
\providecommand \@ifx [1]{%
 \ifx #1\expandafter \@firstoftwo
 \else \expandafter \@secondoftwo
 \fi
}%
\providecommand \natexlab [1]{#1}%
\providecommand \enquote  [1]{``#1''}%
\providecommand \bibnamefont  [1]{#1}%
\providecommand \bibfnamefont [1]{#1}%
\providecommand \citenamefont [1]{#1}%
\providecommand \href@noop [0]{\@secondoftwo}%
\providecommand \href [0]{\begingroup \@sanitize@url \@href}%
\providecommand \@href[1]{\@@startlink{#1}\@@href}%
\providecommand \@@href[1]{\endgroup#1\@@endlink}%
\providecommand \@sanitize@url [0]{\catcode `\\12\catcode `\$12\catcode
  `\&12\catcode `\#12\catcode `\^12\catcode `\_12\catcode `\%12\relax}%
\providecommand \@@startlink[1]{}%
\providecommand \@@endlink[0]{}%
\providecommand \url  [0]{\begingroup\@sanitize@url \@url }%
\providecommand \@url [1]{\endgroup\@href {#1}{\urlprefix }}%
\providecommand \urlprefix  [0]{URL }%
\providecommand \Eprint [0]{\href }%
\providecommand \doibase [0]{http://dx.doi.org/}%
\providecommand \selectlanguage [0]{\@gobble}%
\providecommand \bibinfo  [0]{\@secondoftwo}%
\providecommand \bibfield  [0]{\@secondoftwo}%
\providecommand \translation [1]{[#1]}%
\providecommand \BibitemOpen [0]{}%
\providecommand \bibitemStop [0]{}%
\providecommand \bibitemNoStop [0]{.\EOS\space}%
\providecommand \EOS [0]{\spacefactor3000\relax}%
\providecommand \BibitemShut  [1]{\csname bibitem#1\endcsname}%
\let\auto@bib@innerbib\@empty
%</preamble>
\end{thebibliography}%


\begin{thebibliography}{24}%
\makeatletter
\providecommand \@ifxundefined [1]{%
 \@ifx{#1\undefined}
}%
\providecommand \@ifnum [1]{%
 \ifnum #1\expandafter \@firstoftwo
 \else \expandafter \@secondoftwo
 \fi
}%
\providecommand \@ifx [1]{%
 \ifx #1\expandafter \@firstoftwo
 \else \expandafter \@secondoftwo
 \fi
}%
\providecommand \natexlab [1]{#1}%
\providecommand \enquote  [1]{``#1''}%
\providecommand \bibnamefont  [1]{#1}%
\providecommand \bibfnamefont [1]{#1}%
\providecommand \citenamefont [1]{#1}%
\providecommand \href@noop [0]{\@secondoftwo}%
\providecommand \href [0]{\begingroup \@sanitize@url \@href}%
\providecommand \@href[1]{\@@startlink{#1}\@@href}%
\providecommand \@@href[1]{\endgroup#1\@@endlink}%
\providecommand \@sanitize@url [0]{\catcode `\\12\catcode `\$12\catcode
  `\&12\catcode `\#12\catcode `\^12\catcode `\_12\catcode `\%12\relax}%
\providecommand \@@startlink[1]{}%
\providecommand \@@endlink[0]{}%
\providecommand \url  [0]{\begingroup\@sanitize@url \@url }%
\providecommand \@url [1]{\endgroup\@href {#1}{\urlprefix }}%
\providecommand \urlprefix  [0]{URL }%
\providecommand \Eprint [0]{\href }%
\providecommand \doibase [0]{http://dx.doi.org/}%
\providecommand \selectlanguage [0]{\@gobble}%
\providecommand \bibinfo  [0]{\@secondoftwo}%
\providecommand \bibfield  [0]{\@secondoftwo}%
\providecommand \translation [1]{[#1]}%
\providecommand \BibitemOpen [0]{}%
\providecommand \bibitemStop [0]{}%
\providecommand \bibitemNoStop [0]{.\EOS\space}%
\providecommand \EOS [0]{\spacefactor3000\relax}%
\providecommand \BibitemShut  [1]{\csname bibitem#1\endcsname}%
\let\auto@bib@innerbib\@empty
%</preamble>
\bibitem [{\citenamefont {Binder}\ and\ \citenamefont
  {Heermann}(2010)}]{Binder:1339249}%
  \BibitemOpen
  \bibfield  {author} {\bibinfo {author} {\bibfnamefont {K.}~\bibnamefont
  {Binder}}\ and\ \bibinfo {author} {\bibfnamefont {D.~W.}\ \bibnamefont
  {Heermann}},\ }\href {\doibase 10.1007/978-3-642-03163-2} {\emph {\bibinfo
  {title} {{Monte Carlo Simulation in Statistical Physics: An Introduction; 5th
  ed.}}}},\ Graduate Texts in Physics\ (\bibinfo  {publisher} {Springer},\
  \bibinfo {address} {Berlin, Heidelberg},\ \bibinfo {year} {2010})\BibitemShut
  {NoStop}%
\bibitem [{\citenamefont {Foulkes}\ \emph {et~al.}(2001)\citenamefont
  {Foulkes}, \citenamefont {Mitas}, \citenamefont {Needs},\ and\ \citenamefont
  {Rajagopal}}]{Foulkes2001}%
  \BibitemOpen
  \bibfield  {author} {\bibinfo {author} {\bibfnamefont {W.~M.~C.}\
  \bibnamefont {Foulkes}}, \bibinfo {author} {\bibfnamefont {L.}~\bibnamefont
  {Mitas}}, \bibinfo {author} {\bibfnamefont {R.~J.}\ \bibnamefont {Needs}}, \
  and\ \bibinfo {author} {\bibfnamefont {G.}~\bibnamefont {Rajagopal}},\ }\href
  {\doibase 10.1103/RevModPhys.73.33} {\bibfield  {journal} {\bibinfo
  {journal} {Rev. Mod. Phys.}\ }\textbf {\bibinfo {volume} {73}},\ \bibinfo
  {pages} {33} (\bibinfo {year} {2001})}\BibitemShut {NoStop}%
\bibitem [{\citenamefont {Lynn}\ \emph {et~al.}(2019)\citenamefont {Lynn},
  \citenamefont {Tews}, \citenamefont {Gandolfi},\ and\ \citenamefont
  {Lovato}}]{Lyn2019}%
  \BibitemOpen
  \bibfield  {author} {\bibinfo {author} {\bibfnamefont {J.}~\bibnamefont
  {Lynn}}, \bibinfo {author} {\bibfnamefont {I.}~\bibnamefont {Tews}}, \bibinfo
  {author} {\bibfnamefont {S.}~\bibnamefont {Gandolfi}}, \ and\ \bibinfo
  {author} {\bibfnamefont {A.}~\bibnamefont {Lovato}},\ }\href {\doibase
  10.1146/annurev-nucl-101918-023600} {\bibfield  {journal} {\bibinfo
  {journal} {Annu. Rev. Nucl. Part. Sci.}\ }\textbf {\bibinfo {volume} {69}},\
  \bibinfo {pages} {279} (\bibinfo {year} {2019})}\BibitemShut {NoStop}%
\bibitem [{\citenamefont {Goodman}\ and\ \citenamefont
  {Sokal}(1989)}]{PhysRevD.40.2035}%
  \BibitemOpen
  \bibfield  {author} {\bibinfo {author} {\bibfnamefont {J.}~\bibnamefont
  {Goodman}}\ and\ \bibinfo {author} {\bibfnamefont {A.~D.}\ \bibnamefont
  {Sokal}},\ }\href {\doibase 10.1103/PhysRevD.40.2035} {\bibfield  {journal}
  {\bibinfo  {journal} {Phys. Rev. D}\ }\textbf {\bibinfo {volume} {40}},\
  \bibinfo {pages} {2035} (\bibinfo {year} {1989})}\BibitemShut {NoStop}%
\bibitem [{\citenamefont {Weare}(2007)}]{weare2007}%
  \BibitemOpen
  \bibfield  {author} {\bibinfo {author} {\bibfnamefont {J.}~\bibnamefont
  {Weare}},\ }\href {\doibase 10.1073/pnas.0705418104} {\bibfield  {journal}
  {\bibinfo  {journal} {Proc. Natl. Acad. Sci. U.S.A.}\ }\textbf {\bibinfo
  {volume} {104}},\ \bibinfo {pages} {12657} (\bibinfo {year}
  {2007})}\BibitemShut {NoStop}%
\bibitem [{\citenamefont {Toulouse}\ \emph {et~al.}(2016)\citenamefont
  {Toulouse}, \citenamefont {Assaraf},\ and\ \citenamefont
  {Umrigar}}]{TOULOUSE2016285}%
  \BibitemOpen
  \bibfield  {author} {\bibinfo {author} {\bibfnamefont {J.}~\bibnamefont
  {Toulouse}}, \bibinfo {author} {\bibfnamefont {R.}~\bibnamefont {Assaraf}}, \
  and\ \bibinfo {author} {\bibfnamefont {C.~J.}\ \bibnamefont {Umrigar}},\
  }\href {\doibase http://dx.doi.org/10.1016/bs.aiq.2015.07.003} {\bibfield
  {journal} {\bibinfo  {journal} {Adv. Quantum Chem.}\ }\textbf {\bibinfo
  {volume} {73}},\ \bibinfo {pages} {285 } (\bibinfo {year}
  {2016})}\BibitemShut {NoStop}%
\bibitem [{\citenamefont {Mohr}\ \emph {et~al.}(2018)\citenamefont {Mohr},
  \citenamefont {Eixarch}, \citenamefont {Amsler}, \citenamefont {Mantsinen},\
  and\ \citenamefont {Genovese}}]{MOHR201864}%
  \BibitemOpen
  \bibfield  {author} {\bibinfo {author} {\bibfnamefont {S.}~\bibnamefont
  {Mohr}}, \bibinfo {author} {\bibfnamefont {M.}~\bibnamefont {Eixarch}},
  \bibinfo {author} {\bibfnamefont {M.}~\bibnamefont {Amsler}}, \bibinfo
  {author} {\bibfnamefont {M.~J.}\ \bibnamefont {Mantsinen}}, \ and\ \bibinfo
  {author} {\bibfnamefont {L.}~\bibnamefont {Genovese}},\ }\href {\doibase
  https://doi.org/10.1016/j.nme.2018.01.002} {\bibfield  {journal} {\bibinfo
  {journal} {Nucl. Mater. Energy}\ }\textbf {\bibinfo {volume} {15}},\ \bibinfo
  {pages} {64 } (\bibinfo {year} {2018})}\BibitemShut {NoStop}%
\bibitem [{\citenamefont {Assaraf}\ and\ \citenamefont
  {Caffarel}(1999)}]{assaraf:4682}%
  \BibitemOpen
  \bibfield  {author} {\bibinfo {author} {\bibfnamefont {R.}~\bibnamefont
  {Assaraf}}\ and\ \bibinfo {author} {\bibfnamefont {M.}~\bibnamefont
  {Caffarel}},\ }\href {https://doi.org/10.1103/PhysRevLett.83.4682} {\bibfield
   {journal} {\bibinfo  {journal} {Phys. Rev. Lett.}\ }\textbf {\bibinfo
  {volume} {83}},\ \bibinfo {pages} {4682} (\bibinfo {year}
  {1999})}\BibitemShut {NoStop}%
\bibitem [{\citenamefont {Mira}\ \emph {et~al.}(2012)\citenamefont {Mira},
  \citenamefont {Solgi},\ and\ \citenamefont {Imparato}}]{Mira2012}%
  \BibitemOpen
  \bibfield  {author} {\bibinfo {author} {\bibfnamefont {A.}~\bibnamefont
  {Mira}}, \bibinfo {author} {\bibfnamefont {R.}~\bibnamefont {Solgi}}, \ and\
  \bibinfo {author} {\bibfnamefont {D.}~\bibnamefont {Imparato}},\ }\href
  {\doibase 10.1007/s11222-012-9344-6} {\bibfield  {journal} {\bibinfo
  {journal} {Stat. Comput.}\ }\textbf {\bibinfo {volume} {23}},\ \bibinfo
  {pages} {653} (\bibinfo {year} {2012})}\BibitemShut {NoStop}%
\bibitem [{\citenamefont {Borgis}\ \emph {et~al.}(2013)\citenamefont {Borgis},
  \citenamefont {Assaraf}, \citenamefont {Rotenberg},\ and\ \citenamefont
  {Vuilleumier}}]{Borgis_doi:10.1080/00268976.2013.838316}%
  \BibitemOpen
  \bibfield  {author} {\bibinfo {author} {\bibfnamefont {D.}~\bibnamefont
  {Borgis}}, \bibinfo {author} {\bibfnamefont {R.}~\bibnamefont {Assaraf}},
  \bibinfo {author} {\bibfnamefont {B.}~\bibnamefont {Rotenberg}}, \ and\
  \bibinfo {author} {\bibfnamefont {R.}~\bibnamefont {Vuilleumier}},\ }\href
  {\doibase 10.1080/00268976.2013.838316} {\bibfield  {journal} {\bibinfo
  {journal} {Mol. Phys.}\ }\textbf {\bibinfo {volume} {111}},\ \bibinfo {pages}
  {3486} (\bibinfo {year} {2013})}\BibitemShut {NoStop}%
\bibitem [{\citenamefont {Assaraf}\ and\ \citenamefont
  {Domin}(2014)}]{assaraf_PhysRevE.89.033304}%
  \BibitemOpen
  \bibfield  {author} {\bibinfo {author} {\bibfnamefont {R.}~\bibnamefont
  {Assaraf}}\ and\ \bibinfo {author} {\bibfnamefont {D.}~\bibnamefont
  {Domin}},\ }\href {\doibase 10.1103/PhysRevE.89.033304} {\bibfield  {journal}
  {\bibinfo  {journal} {Phys. Rev. E}\ }\textbf {\bibinfo {volume} {89}},\
  \bibinfo {pages} {033304} (\bibinfo {year} {2014})}\BibitemShut {NoStop}%
\bibitem [{\citenamefont {Williamson}\ \emph {et~al.}(2001)\citenamefont
  {Williamson}, \citenamefont {Hood},\ and\ \citenamefont
  {Grossman}}]{PhysRevLett.87.246406}%
  \BibitemOpen
  \bibfield  {author} {\bibinfo {author} {\bibfnamefont {A.~J.}\ \bibnamefont
  {Williamson}}, \bibinfo {author} {\bibfnamefont {R.~Q.}\ \bibnamefont
  {Hood}}, \ and\ \bibinfo {author} {\bibfnamefont {J.~C.}\ \bibnamefont
  {Grossman}},\ }\href {\doibase 10.1103/PhysRevLett.87.246406} {\bibfield
  {journal} {\bibinfo  {journal} {Phys. Rev. Lett.}\ }\textbf {\bibinfo
  {volume} {87}},\ \bibinfo {pages} {246406} (\bibinfo {year}
  {2001})}\BibitemShut {NoStop}%
\bibitem [{\citenamefont {Assaraf}(2014)}]{PhysRevE.90.063317}%
  \BibitemOpen
  \bibfield  {author} {\bibinfo {author} {\bibfnamefont {R.}~\bibnamefont
  {Assaraf}},\ }\href {\doibase 10.1103/PhysRevE.90.063317} {\bibfield
  {journal} {\bibinfo  {journal} {Phys. Rev. E}\ }\textbf {\bibinfo {volume}
  {90}},\ \bibinfo {pages} {063317} (\bibinfo {year} {2014})}\BibitemShut
  {NoStop}%
\bibitem [{\citenamefont {Assaraf}\ \emph
  {et~al.}(2017{\natexlab{a}})\citenamefont {Assaraf}, \citenamefont
  {Jourdain}, \citenamefont {Leli{\`e}vre},\ and\ \citenamefont
  {Roux}}]{Assaraf2017}%
  \BibitemOpen
  \bibfield  {author} {\bibinfo {author} {\bibfnamefont {R.}~\bibnamefont
  {Assaraf}}, \bibinfo {author} {\bibfnamefont {B.}~\bibnamefont {Jourdain}},
  \bibinfo {author} {\bibfnamefont {T.}~\bibnamefont {Leli{\`e}vre}}, \ and\
  \bibinfo {author} {\bibfnamefont {R.}~\bibnamefont {Roux}},\ }\href {\doibase
  10.1007/s40072-017-0105-6} {\bibfield  {journal} {\bibinfo  {journal} {Stoch.
  Partial Differ. Equ.: Anal. Comput.}\ }\textbf {\bibinfo {volume} {6}},\
  \bibinfo {pages} {125} (\bibinfo {year} {2017}{\natexlab{a}})}\BibitemShut
  {NoStop}%
\bibitem [{\citenamefont {Cohen}\ \emph {et~al.}(2015)\citenamefont {Cohen},
  \citenamefont {Gull}, \citenamefont {Reichman},\ and\ \citenamefont
  {Millis}}]{Cohen_2015}%
  \BibitemOpen
  \bibfield  {author} {\bibinfo {author} {\bibfnamefont {G.}~\bibnamefont
  {Cohen}}, \bibinfo {author} {\bibfnamefont {E.}~\bibnamefont {Gull}},
  \bibinfo {author} {\bibfnamefont {D.~R.}\ \bibnamefont {Reichman}}, \ and\
  \bibinfo {author} {\bibfnamefont {A.~J.}\ \bibnamefont {Millis}},\ }\href
  {\doibase 10.1103/physrevlett.115.266802} {\bibfield  {journal} {\bibinfo
  {journal} {Phys. Rev. Lett.}\ }\textbf {\bibinfo {volume} {115}},\ \bibinfo
  {pages} {266802} (\bibinfo {year} {2015})}\BibitemShut {NoStop}%
\bibitem [{\citenamefont {Feldt}\ and\ \citenamefont
  {Assaraf}(2021)}]{doi:10.1021/acs.jctc.0c01069}%
  \BibitemOpen
  \bibfield  {author} {\bibinfo {author} {\bibfnamefont {J.}~\bibnamefont
  {Feldt}}\ and\ \bibinfo {author} {\bibfnamefont {R.}~\bibnamefont
  {Assaraf}},\ }\href {\doibase 10.1021/acs.jctc.0c01069} {\bibfield  {journal}
  {\bibinfo  {journal} {J. Chem. Theory Comput.}\ }\textbf {\bibinfo {volume}
  {17}},\ \bibinfo {pages} {1380} (\bibinfo {year} {2021})}\BibitemShut
  {NoStop}%
\bibitem [{\citenamefont {White}(1992)}]{Wite1992}%
  \BibitemOpen
  \bibfield  {author} {\bibinfo {author} {\bibfnamefont {S.~R.}\ \bibnamefont
  {White}},\ }\href {\doibase 10.1103/PhysRevLett.69.2863} {\bibfield
  {journal} {\bibinfo  {journal} {Phys. Rev. Lett.}\ }\textbf {\bibinfo
  {volume} {69}},\ \bibinfo {pages} {2863} (\bibinfo {year}
  {1992})}\BibitemShut {NoStop}%
\bibitem [{\citenamefont {Knizia}\ and\ \citenamefont
  {Chan}(2012)}]{KniCha-PRL-12}%
  \BibitemOpen
  \bibfield  {author} {\bibinfo {author} {\bibfnamefont {G.}~\bibnamefont
  {Knizia}}\ and\ \bibinfo {author} {\bibfnamefont {G.~K.-L.}\ \bibnamefont
  {Chan}},\ }\href {https://doi.org/10.1103/PhysRevLett.109.186404} {\bibfield
  {journal} {\bibinfo  {journal} {Phys. Rev. Lett.}\ }\textbf {\bibinfo
  {volume} {109}},\ \bibinfo {pages} {186404} (\bibinfo {year}
  {2012})}\BibitemShut {NoStop}%
\bibitem [{\citenamefont {Zahariev}\ and\ \citenamefont
  {Gordon}(2021)}]{Zahariev2021CombinedQM}%
  \BibitemOpen
  \bibfield  {author} {\bibinfo {author} {\bibfnamefont {F.}~\bibnamefont
  {Zahariev}}\ and\ \bibinfo {author} {\bibfnamefont {M.~S.}\ \bibnamefont
  {Gordon}},\ }\href {https://doi.org/10.1039/D0CP06528E} {\bibfield  {journal}
  {\bibinfo  {journal} {Phys. Chem. Chem. Phys.}\ }\textbf {\bibinfo {volume}
  {23}},\ \bibinfo {pages} {14308} (\bibinfo {year} {2021})}\BibitemShut
  {NoStop}%
\bibitem [{\citenamefont {Filippi}\ \emph {et~al.}(2016)\citenamefont
  {Filippi}, \citenamefont {Assaraf},\ and\ \citenamefont
  {Moroni}}]{doi:10.1063/1.4948778}%
  \BibitemOpen
  \bibfield  {author} {\bibinfo {author} {\bibfnamefont {C.}~\bibnamefont
  {Filippi}}, \bibinfo {author} {\bibfnamefont {R.}~\bibnamefont {Assaraf}}, \
  and\ \bibinfo {author} {\bibfnamefont {S.}~\bibnamefont {Moroni}},\ }\href
  {\doibase 10.1063/1.4948778} {\bibfield  {journal} {\bibinfo  {journal} {J.
  Chem. Phys.}\ }\textbf {\bibinfo {volume} {144}},\ \bibinfo {pages} {194105}
  (\bibinfo {year} {2016})}\BibitemShut {NoStop}%
\bibitem [{\citenamefont {Assaraf}\ \emph
  {et~al.}(2017{\natexlab{b}})\citenamefont {Assaraf}, \citenamefont {Moroni},\
  and\ \citenamefont {Filippi}}]{slater_2017}%
  \BibitemOpen
  \bibfield  {author} {\bibinfo {author} {\bibfnamefont {R.}~\bibnamefont
  {Assaraf}}, \bibinfo {author} {\bibfnamefont {S.}~\bibnamefont {Moroni}}, \
  and\ \bibinfo {author} {\bibfnamefont {C.}~\bibnamefont {Filippi}},\ }\href
  {\doibase 10.1021/acs.jctc.7b00648} {\bibfield  {journal} {\bibinfo
  {journal} {J. Chem. Theory Comput.}\ }\textbf {\bibinfo {volume} {13}},\
  \bibinfo {pages} {5273} (\bibinfo {year} {2017}{\natexlab{b}})}\BibitemShut
  {NoStop}%
\bibitem [{\citenamefont {Cyrot}(1977)}]{cyrot1977hubbard}%
  \BibitemOpen
  \bibfield  {author} {\bibinfo {author} {\bibfnamefont {M.}~\bibnamefont
  {Cyrot}},\ }\href@noop {} {\bibfield  {journal} {\bibinfo  {journal} {Physica
  B+C}\ }\textbf {\bibinfo {volume} {91}},\ \bibinfo {pages} {141} (\bibinfo
  {year} {1977})}\BibitemShut {NoStop}%
\bibitem [{\citenamefont {Baroni}\ and\ \citenamefont
  {Moroni}(1999)}]{Baroni1999ReptationQM}%
  \BibitemOpen
  \bibfield  {author} {\bibinfo {author} {\bibfnamefont {S.}~\bibnamefont
  {Baroni}}\ and\ \bibinfo {author} {\bibfnamefont {S.}~\bibnamefont
  {Moroni}},\ }\href {https://doi.org/10.1103/PhysRevLett.82.4745} {\bibfield
  {journal} {\bibinfo  {journal} {Phys. Rev. Lett.}\ }\textbf {\bibinfo
  {volume} {82}},\ \bibinfo {pages} {4745} (\bibinfo {year}
  {1999})}\BibitemShut {NoStop}%
\bibitem [{\citenamefont {Shumway}\ and\ \citenamefont
  {Gilbert}(2006)}]{Shumway2006}%
  \BibitemOpen
  \bibfield  {author} {\bibinfo {author} {\bibfnamefont {J.}~\bibnamefont
  {Shumway}}\ and\ \bibinfo {author} {\bibfnamefont {M.}~\bibnamefont
  {Gilbert}},\ }\href {\doibase 10.1088/1742-6596/35/1/017} {\bibfield
  {journal} {\bibinfo  {journal} {J. Phys. Conf. Ser.}\ }\textbf {\bibinfo
  {volume} {35}},\ \bibinfo {pages} {190} (\bibinfo {year} {2006})}\BibitemShut
  {NoStop}%
\end{thebibliography}
\end{document}